\documentclass[fleqn,usenatbib,referee]{aa}

\usepackage{graphicx}
\usepackage{txfonts}
\usepackage{amssymb}	

\def\aap{A\&A}
\def\apjs{ApJS}

\begin{document} 

\title{Lyman and Balmer breaks reveal mature z=8 galaxies with the code P\'egase.3}

   \subtitle{}

   \author{Brigitte Rocca-Volmerange
          \inst{1,2}
          \and
          Antoine Hilberer
          \inst{2}
          \and
          Michel Fioc
          \inst{1}
          }

   \institute{Sorbonne Universit\'e, CNRS, Institut Astrophysique de Paris, 98 bis bd Arago 75014 Paris, France\\
             \email{rocca-volmerange@iap.fr}
                    \and
             Universit\'e Paris-Sud F-91405, Orsay, France\\
    }          
\titlerunning {Mature z=8 galaxies }
\authorrunning {Rocca-Volmerange et al.}
  \date{Received ; accepted}

 
  \abstract
   {In the present paper, we derive the main properties of $z\simeq$8 galaxies as redshift, age, star formation history (SFH) and masses with the help of the new spectro-chemical evolutionary code P\'egase.3 with dust modeling. The attenuation to emission radiative transfer respects the energy balance with the help of MonteCarlo simulations computed for spiral and spheroidal geometries. The galaxy sample (EGSY-2008532660, EGS-zs8-1, EGS-zs8-2), observed with HST and Spitzer/IRAC, is selected for its atypical  [3.6]-[4.5]$\mu$m IR excess. After best-fitting observations with multiple libraries of continuous UV-to-farIR Spectral Energy Distributions (SEDs), the main results are: i) the IR  excess is due to the Balmer break of red giant stars with galaxy ages of 450-500 Myrs  ii) masses are high: 7 $\times$10$^{10}$ M$_\odot$ for stellar mass,  10$^{11}$ M$_\odot$ for galaxy mass, 3 $\times$10$^8$ M$_\odot$ for dust mass iii) they are probing mature (no starbursting) galaxies only with the simulations of spheroidal (and no spiral) geometries and the star formation history (SFH ) predicting  local SEDs and colors of massive ellipticals. Less constrained by observations not covering the Balmer zone, the galaxy GN-z11 is consistent with ages of 160 Myrs, galaxy mass of  6.5 $\times$10$^{10}$ M$_\odot$ at $z\simeq$11-13. These results are robust by confirming the  agreement of the Lyman/Balmer photometric and spectroscopic redshifts as  the low (or null) Ly$\alpha$ emission line, as waited for a resonance line in a metal-enriched medium. These results constrain the rapid and efficient formation of massive elliptical galaxies, significantly distinct of early-type disk-spiral galaxies,  as the reionization of the Universe. Other constraints on mass growth physics, star formation triggering and the link with galaxy-active nucleus are waited for from the improved spatial and angular resolution images, spectra and colors of the future telescopes:  JWST followed by ELT and EUCLID .
    }
 \maketitle
 
%

\section{Introduction}

  Among the most fascinating discoveries  on the last thirty years, the detection of galaxies from z=1 to z$\geq$8, when the universe was less than 1Gyr old, strongly constrains the basic physics of primeval galaxies to the reionization of the universe as well as the establishment of the large scale structures and dark matter haloes. 
At such extreme distances, the determination of redshift is essential. In the best cases, emission lines of galaxies are used, assuming galaxies are young enough to intensively  form stars  with ionizing photons inducing huge (as $H{\alpha}$, [O III])  emission lines. As an example \citet{RobertsBorsani2016}  conclude  that z=8 galaxies  are early star-forming with young (5 Myr) stellar populations  and  very strong equivalent widths ( EW[O III]=1500$\AA$) of nebular line [O III] and a flat continuum. As a consequence, the authors, after deducing a precise redshift, conclude that the [3.6]-[4.5] color excess of their sample is  due to the impact of the [OIII] line on the 4.5$\mu$m band and no comparably bright nebular line in the 3.6 $\mu$m. But estimates of  equivalent widths require non only the  fluxes of emission lines but the attenuated continuum  in the  far-UV to the near-infrared, so a coherent dust modeling is essential.  

When emission lines are not identified, photometric redshifts are estimated on the basis of photometric signatures as the discontinuity of Lyman break, used for the so-called Lyman-Break  Galaxies (LBGs) (e.g., \citet{Steidel1996}, \citet{Steidel2003}). Another signature is  the 4000$\AA$\ discontinuity  (so called D4000), tracer of ages for evolved populations. Due to the red giant spectra, the amplitude of this spectral feature is used to measure the past star formation rates of evolved galaxies, in particular for early-type elliptical galaxies, \citet{Bruzual1983}. Moreover when a single stellar population (SSP) evolves, massive UV-luminous stars (of short lifetimes) rapidly disappear by exploding as supernovae: the amplitude of the Lyman break as  the far-UV continuum slope rapidly decrease.  As a consequence, the corresponding spectrum  evolution of an SSP shows an increasing Balmer discontinuity in coherence with a decreasing  Lyman discontinuity. This remark is however modulated by the star formation history SFH (t), a sum of SSPs of various amplitude and metallicity with dust.  To summarize the simultaneous detection of the Lyman and Balmer breaks not only allows a photometric redshift determination but more their respective amplitudes, not simultaneously varying, are a strong constraint to estimate  age and star formation history. Moreover if the photometric and spectroscopic redshifts correspond, they make more robust the synthetic SED templates used to photometric analyzes: this remark is important because the photometric determination being the only way in the deepest surveys, even if limited by the low angular resolution or the dust contribution (attenuation or emission), the survey analyzes with dust modeling codes as P\'egase.3  will be more robust.  

When massive stars explode, their ejecta increase the interstellar metallicity and dust grain mass, and so the dust UV attenuation. Recent articles underline the necessity of following dust mass evolution (\citet{Laporte2017}, \citet{Aravena2016}) and propose estimates at z$\geq$5, \citet{Hirashita2017}, from ALMA limits, associated with UV-optical SEDs giving a very low value M$_{dust}\leq$ 2 $\times$10$^6$M$\odot$, justified by reverse shocks with arbitrarily a faint role of SNe as dust sources. In other models, the UV-to-IR relation  is given by the so-called IRX-$\beta$\ relation from attenuation curves not specifically adapted to $z=8$ galaxies, as the code CIGALE limited at $z=2$ by \citet{LoFaro2017}. As a consequence, in manycases the high-$z$\ effects are not in coherence with other fundamental parameters as masses of stars, dust and metallicity Z see \citet{Laskar2011}, \citet{Maiolino2015} and others.
Once more UV dust extinction is discussed, specifically for LBG by contradictory papers as \citet{Ouchi2004}, \citet{Iwata2007}, \citet{Reddy2008}. In other papers LBG are assimilated with Ly$\alpha$ galaxies, even if at z$\geq$7 only 20$\%$  - 30 $\%$ of faint galaxies show Ly$\alpha$ emission,  \citet{Finkelstein2013}, or with galaxies not detected by ALMA see \citet{Yajima2014}. Last but not least, lensing may also affect the interpretations of stellar masses as photometric redshifts. So that evolution with dust modeling is clearly needed to calibrate the Lyman break amplitude.

Following the previous version P\'egase.2 with metallicity effects, \citet{Fioc1999a}, the present version of the evolutionary code P\'egase.3 (Fioc \& Rocca-Volmerange, submitted) estimates the attenuation and adds the Monte Carlo simulations radiative transfer to dust emission, in coherency with attenuation by metal enrichment. This new model allow to build continuous far-UV-far IR SED librairies from star, gas and dust evolution modeling with MonteCarlo simulations for two  spiral and spheroidal geometries; the case of instantaneous starburst (SB) is also considered. The paper presents the observations of the analyzed $z\simeq$8 galaxy sample selected for the  [3.6]-[4.5]$\mu$m IR excess from the  HST to Spitzer/CANDELS data  in section 2. Redshifts  are determined from Lyman and Balmer break signatures in section 3.  In section 4, we show for the galaxy sample the best-fits of observations and the corresponding outputs as ages, masses and metallicity, the case of GN-z11 is considered. Discussion and conclusion are in section 5.

\section{Observations}
The so-called "Super-Eights" galaxies are selected from the full wide-area Spitzer/CANDELS program, the HST/WFC3  and when possible CFHT, Subaru, Suprime-Cam, VLT and VISTA, see more details in \citet{RobertsBorsani2016}, \citet{Bouwens2015}, \citet{McCracken2012}. The main selection criteria are the intense brightness with Spitzer/CANDELS reaching 25.5 mag in the 3.6 $\mu$m band and 25.3 mag in the 4.5 $\mu$m band.  The Spitzer observations are corrected from the 2 arcsec to total aperture from individual profiles plus PSF.  The three galaxies of the sample are EGSY-2008532660 (hereafter EGSY), EGS-zs8-1 (EGS1), EGS-zs8-2 (EGS2), all with already defined IR excess. For EGSY, the magnitude H$_{AB}$ is 25.26 with the photometric redshift  z= 8.57$^{+0.22}_{-0.43}$. Observed with the MOSFIRE spectrograph at the Keck Observatory, the Lyman$\alpha$ (Ly$\alpha$) emission line has been discovered at the spectroscopic redshift 8.683$^{+0.001}_{-0.004}$, a value coherent with its photometric redshift,  \citet{Zitrin2015}. Its rest-frame equivalent width EW(Ly$\alpha$) is from 17 to 42$\AA$ .The other galaxy EGS-zs8-1 is bright in the rest-frame UV with M$_{UV}$= -21.2. Its Ly$\alpha$ emission also discovered with MOSFIRE, \citet{Song2016}, confirms a redshift of 7.6637 $\pm$ 0.0011 and a low rest-frame equivalent width of EW(Ly$\alpha$)= 15.6$^{+5.9}_{-3.6} \AA$. The third galaxy EGS-zs8-2  has only a tentative Lyman$\alpha$ emission feature at a possible value of $z$=7.47.These faint values of EW(Ly$\alpha$) associated  to the non-detection of the line in the other 11 $z\simeq$ 7-8 galaxies imply the faint contribution of this resonance line, highly sensitive to the metal content Z. For this reason, reporting below the discussion  on the line emissions line with  metallicity Z, we limit our analysis to the attenuated stellar and nebular continua with metallicity.    
\begin{table}
\begin{footnotesize}
\begin{center}
\begin{tabular}{c|c|c|c|c|c|c|c|c|c|c|c}
  Galaxie / $\lambda_{filter} (\mu$m) & 0.38 & 0.49 & 0.59 & 0.77 & 0.88 & 1.06 & 1.25 & 1.40 & 1.54 & 2.15\\
  \hline
  EGSY-2008532660	 & 16.5 & <$10^{-3}$ & <$10^{-3}$ & <$10^{-3}$ & 19.6 & - & 220.0 & 283.5 & 285.9 & 311.8\\
  \hline
  EGS-zs8-1 & <$10^{-3}$ & <$10^{-3}$ & - & 27.4 & 14.7 & 126.3 & 315.8 & - & 364.7 & 190.5\\
  \hline
  EGS-zs8-2 & <$10^{-3}$ & <$10^{-3}$ & 0.1 & - & <$10^{-3}$ & 170.7 & 314.3 & - & 325.6 & 497.3\\
  \hline
GN-z11& - & 7 $\pm$ 9 & 2 $\pm$7& 5$\pm$10 & 17$\pm$11 & -7$\pm$ 9& 11$\pm$8 & 64$\pm$13 &152$\pm$10&137$\pm$67\\
\end{tabular}
\\[0.5cm]%
\begin{tabular}{c|c|c|c}
  Galaxie / $\lambda_{filter} (\mu$m) & 3.6 & 4.5\\
  \hline
  EGSY-2008532660	 & 398.2 & 805.9 \\
  \hline
  EGS-zs8-1 & 498.6 & 800.0\\
  \hline
  EGS-zs8-2 & 341.3 & 831.7\\
  \hline
 GN-z11& 139$\pm$21 &  144$\pm$27\\
  \end{tabular}
\end{center}
\end{footnotesize}
\caption {\textit{The galaxy sample: photometric fluxes of galaxies are given in nJy by \citet{RobertsBorsani2016}. The authors give error bars for the z$\simeq$8 galaxy sample and \citet{Oesch2016} on its table 4 for GN-z11. The  effective wavelength $\lambda_{filter}$ of the HST F606W, F814W, F105W, F125W, F140W, et F160W filters are given as the  3.6 and 4.5$\mu$m from Spitzer/IRAC and the 2.15$\mu$m of the filter Ks from Subaru/CIAO.}}
\end{table}

\section {Fluxes and photometric redshifts}
The apparent fluxes of the galaxy sample are corrected for luminosity distance $D_L$,  according to the following equations from \citet{Hogg1999}.  
 \begin{equation}
F_\lambda^{receipted}=\frac{1}{1+z}\frac{L_{\lambda/(1+z)}^{emitted}}{4\pi D_L^2}
 \label{DL}
\end{equation}
\begin{equation}
\begin{aligned}
D_L&=\frac{c}{H_0}(1+z) \bigg[ \eta(1,\Omega_0)-\eta(\frac{1}{1+z},\Omega_0)\bigg] \\
\eta(a,\Omega_0)&=2\sqrt{s^3+1}\bigg[\frac{1}{a^4}-0.1540\frac{s}{a^3}+0.4304\frac{s^2}{a^2}+0.19097\frac{s^3}{a}+0.066941s^4 \bigg]^{-1/8} \\
s^3&=\frac{1-\Omega_0}{\Omega_0}
\end{aligned}
\end{equation}
where $H_0=70 km.s^{-1}.Mpc^{-1}$  and $\Omega_0$=0.3  where $\Omega_0.$ is the density parameter of the baryonic  mass and dark energy. 

At such distances, spectroscopic redshifts could be derived from emission lines as Ly$\alpha$ line, [O III] and H$\beta$ or others.As already mentioned,these redshift estimates were proposed by \citet{RobertsBorsani2016}, \citet{Bouwens2015}.   Photometric redshifts $z_{LyB}$\ may also be derived from the two strong signatures which are the Lyman break (912$\AA$) and the Balmer break ($\simeq4000\AA$), both covered by the HST and Spitzer (optical to mid-IR) observations of our sample. The objectives are to compare spectroscopic and photometric redshifts,  the robustness of the photometric redshift is then confirmed. The first example is proposed for the  galaxy EGSY-2008532660. As shown on Fig 1, the distance of the two  identified breaks (Lyman and Balmer) identifies the minimization process from the best-fitted SED shape within $\pm10\%$\  the photometric redshift $z_{LyB}$=8.62$^{+1.6}_{-0.7}$, remarquably consistent with the spectroscopic redshift $z_{spec}$=8.683$^{+0.001}_{-0.004}$. 
\begin{figure}
\begin{center}
\includegraphics[width=12cm,] {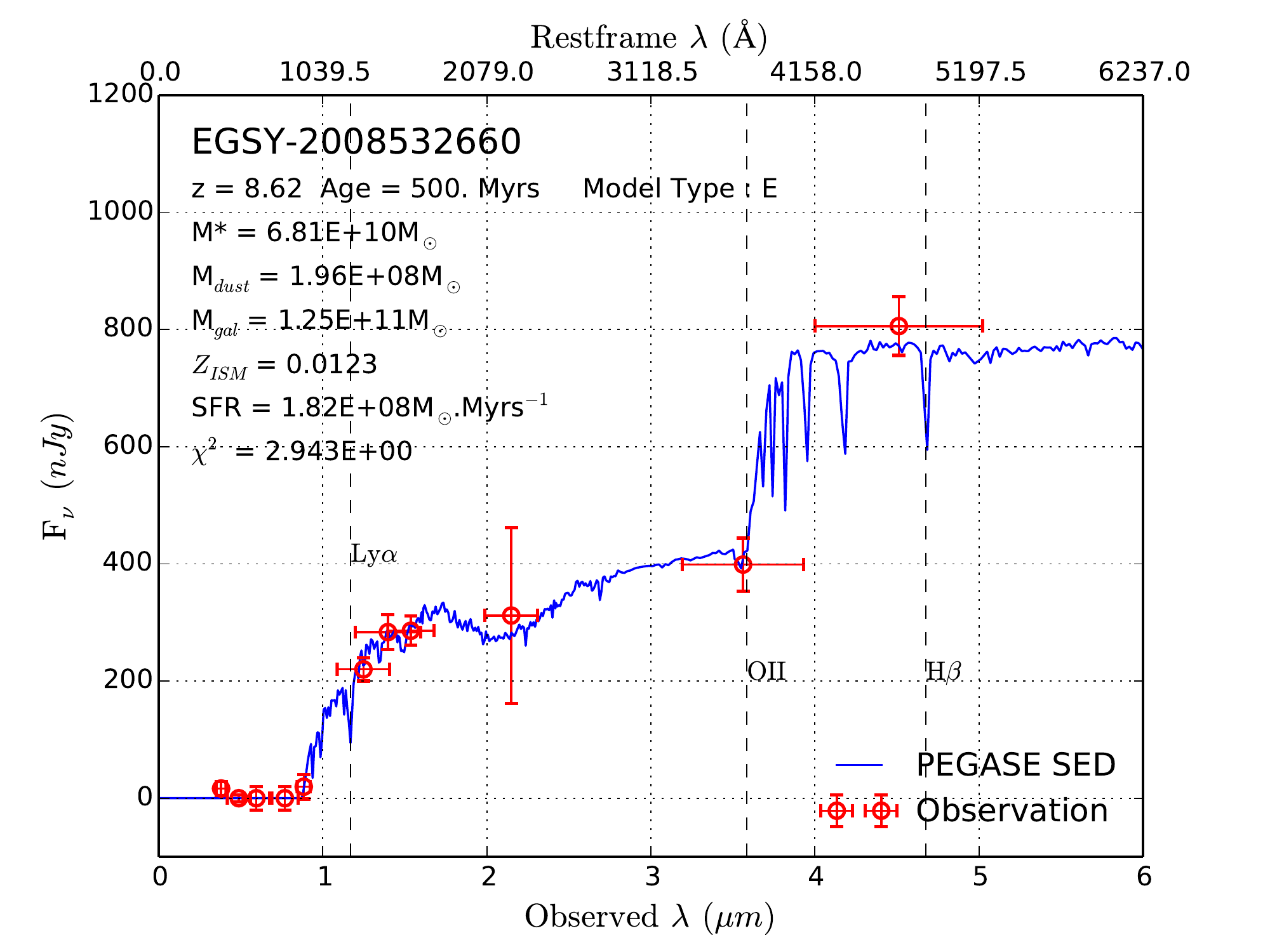}
\includegraphics[width=6cm, angle=0] {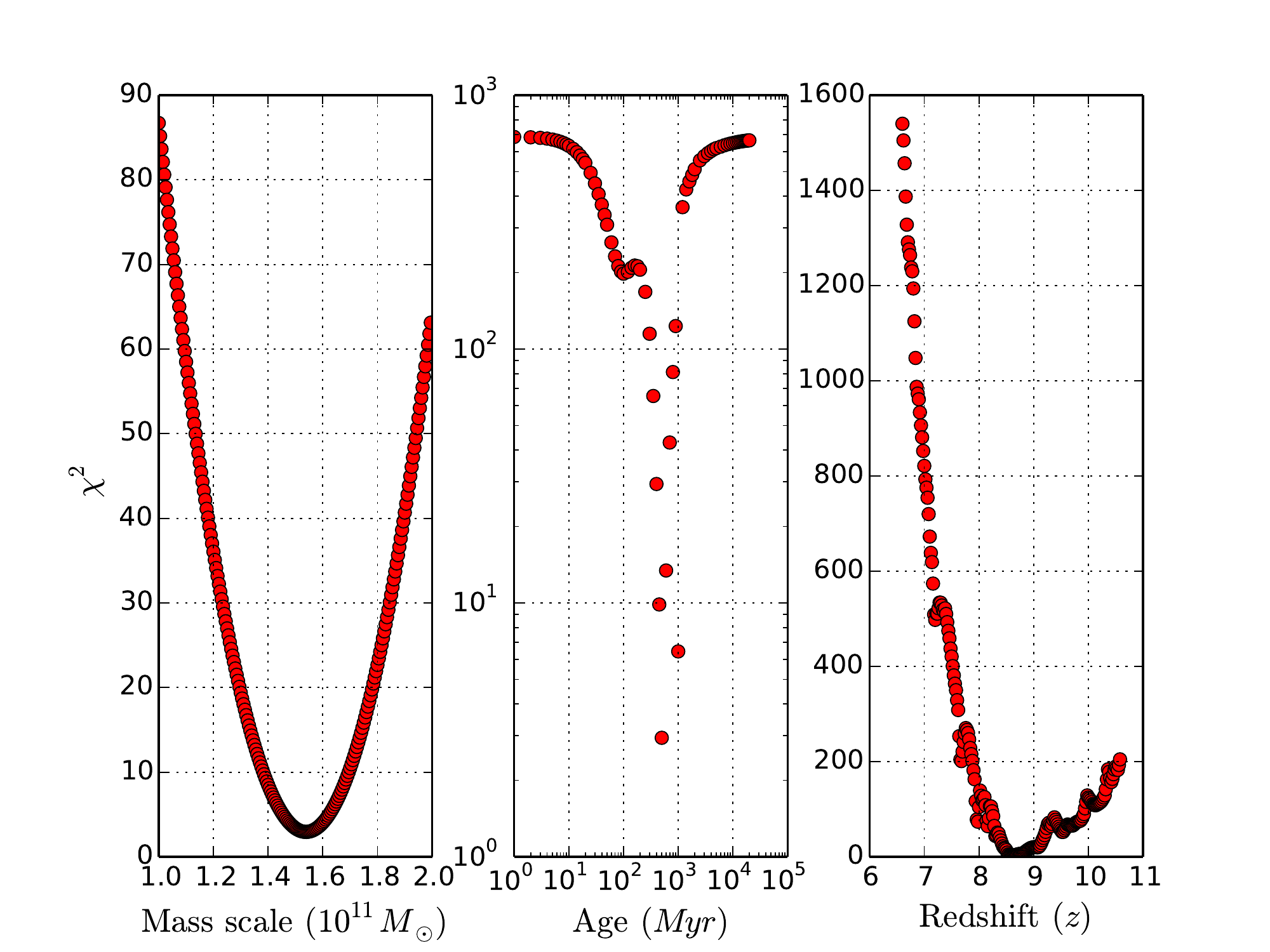}    
\caption{\textit{Redshift confirmation at z=8.6 of the  galaxy EGSY-2008532660 determined from the Lyman  ($912\AA$)  and Balmer ($\simeq 4000\AA$) breaks. The other properties are discussed in the text. Top: The best fitted SED (blue line) is compared with observations (red crosses for apparent fluxes and error bars). Bottom: The $\chi^2$ evolution and minimization for ages, masses and redshifts.  }}
\end{center}
\label{EGSYE}
 \end{figure}

  \section{Best-fits with galaxy templates}  
 \subsection { A brief recall of P\'egase.3} 
 The code  predicts for a large variety of proposed star formation scenarios (star formation laws, inflow and outflow rates, IMF)  the continuous UV to far-IR spectral energy distributions (SEDs) and colors by following metal-dependent evolution of stars, gas, abundances and dust modeling by radiative transfer Monte Carlo simulations. SED templates and many outputs, taking into account coherent attenuation and emission by grains,  are computed on 20Gyrs ages with the resolution of 1Myr. Colors are computed for a large variety of instruments and telescopes maybe extended to the future generations of instruments, in particular for JWST and EUCLID. The system is the galaxy itself with exchanges by inflows from reservoirs of by outflows towards the extragalactic medium. The interstellar medium (ISM) where the chemical evolution is followed is divided in \emph{the diffuse ISM} (DISM) and  compact \emph{star-forming clouds}. Ejecta of the Core Collapse supernovae explosions (CCSNe) as their final remnants (neutron star or stellar black hole: NSBH) are cumulated in the ISM for each scenario increasing respectively the metal-enrichment and the global mass of stellar dense residues, \citet{RoccaVolmerange2015}. Dust attenuation and emission are simulated with properties of classical Draine dust grains distributed in the ISM and heated by the average current radiation field. The Monte Carlo radiative transfer from dust attenuation to emission is computed for two  spiral and spheroidal geometries, respectively assimilated with Spiral Sc and early elliptical (E) types. All details of the code are given in the article (Fioc and Rocca-Volmerange, 2018, submitted ) with a  refined documentation on the P\'egase site (www2.iap.fr/pegase). The authors validate the spiral model by comparing the SED shape, abundances and other outputs with  the Milky Way properties. The spheroidal scenario was also validated for local elliptical galaxies after comparison with the $0\leq z\leq 0.1$\ SDSS galaxies,  as proposed in the color-color diagram by \citet{Tsalmantza2012}. Moreover the scenario of spheroidal galaxies was used at high redshifts to interpret observed SEDs of radiogalaxy hosts, \citet{RoccaVolmerange2013}. Concerning masses, as shown below, $M_{stellar}(t) $ at age t is derived after a best-fit procedure of the normalized SEDs on observations, respecting the constant mass of the system $M_{system}= M_{galaxy}(0) + \sum_{i=1}^{n_{\text{res}}} M_{res, i}(0)$  where the total number of reservoirs is $n_{res}$. A long list of various \emph{initial mass function}s (IMFs) is proposed. 
 
The adopted libraries of templates are respectively for elliptic/spheroidal and spiral galaxies. The instantaneous starburst SB is also used for building a third template library. 
 Taking advantage of the large multi wavelength coverage of observations, continuously extended from the far-UV (HST) to the mid-infrared (Spitzer), we prefer the basic $\chi^2$ minimization on the totality of the wavelength coverage of observational data.  For each observational flux $F_i^ {obs}$ the distance to the synthetic flux $F_i^ {synth}$, after the normalization by a scale-factor $ \alpha$ \, is computed,  the error factor $\sigma_i$ for each point, is assumed gaussian  and the scale factor, also called normalization factor, is unique and identically applied to all $F_i^ {obs}$.  
 \begin{equation}
\chi^2 = \sum^N_{i=1} \bigg( \frac{F^{obs}_i - \alpha \times F^{synth}_i}{\sigma_i} \bigg)^2
\label{khi2}
\end{equation}
 
%
 
\subsection{The galaxy EGSY-2008532660}

 Fig.1
 shows the best-fits of the galaxy EGSY-2008532660 and its $\chi^2$\ minimization curves for the three parameters (mass, age,z) with the help of the spheroidal/elliptical scenario (E) templates with the star formation history of  $\simeq$1Gyr time-scale (see Fig~\ref{SFRz})built to fit the local SEDs. The most surprising result of this galaxy best-fit at the confirmed redshift is the high age of 500Myrs, mainly imposed  by the amplitude of the spectral Balmer discontinuity (D4000), calibrated by the age and number of the giant stars. Moreover this age is coherent  with the far-UV data fitted by the dust attenuated slope of massive but evolved stars. The current star formation rate (180 $M_{\odot}$ by year) is active, while the cumulated star formation history was particularly intense when compared to other scenarios. While   the comparison with the SFH of an  instantaneous starburst (SB) is proposed below, the spiral scenario predicts SEDs unable to fit observations, whatever ages. The best-fits fully reproduce the [3.6]-[4.5] $\mu$m IR excess with a population of evolved giant stars, and calibrated on observations, give respective masses of 6.8 $\times $10$^{10}$ M$_{\odot}$ for live stars (the stellar mass) and 2 $\times $10$^{8}$ M$_{\odot}$ for dust and 10$^{11}$ M$_{\odot}$ for the global galaxy. Other outputs as metallicity are presented on Table~\ref{Results}.

\subsection{The galaxies EGS2 and EGS1}
\begin{figure}
\begin{center}
\includegraphics[width=12cm] {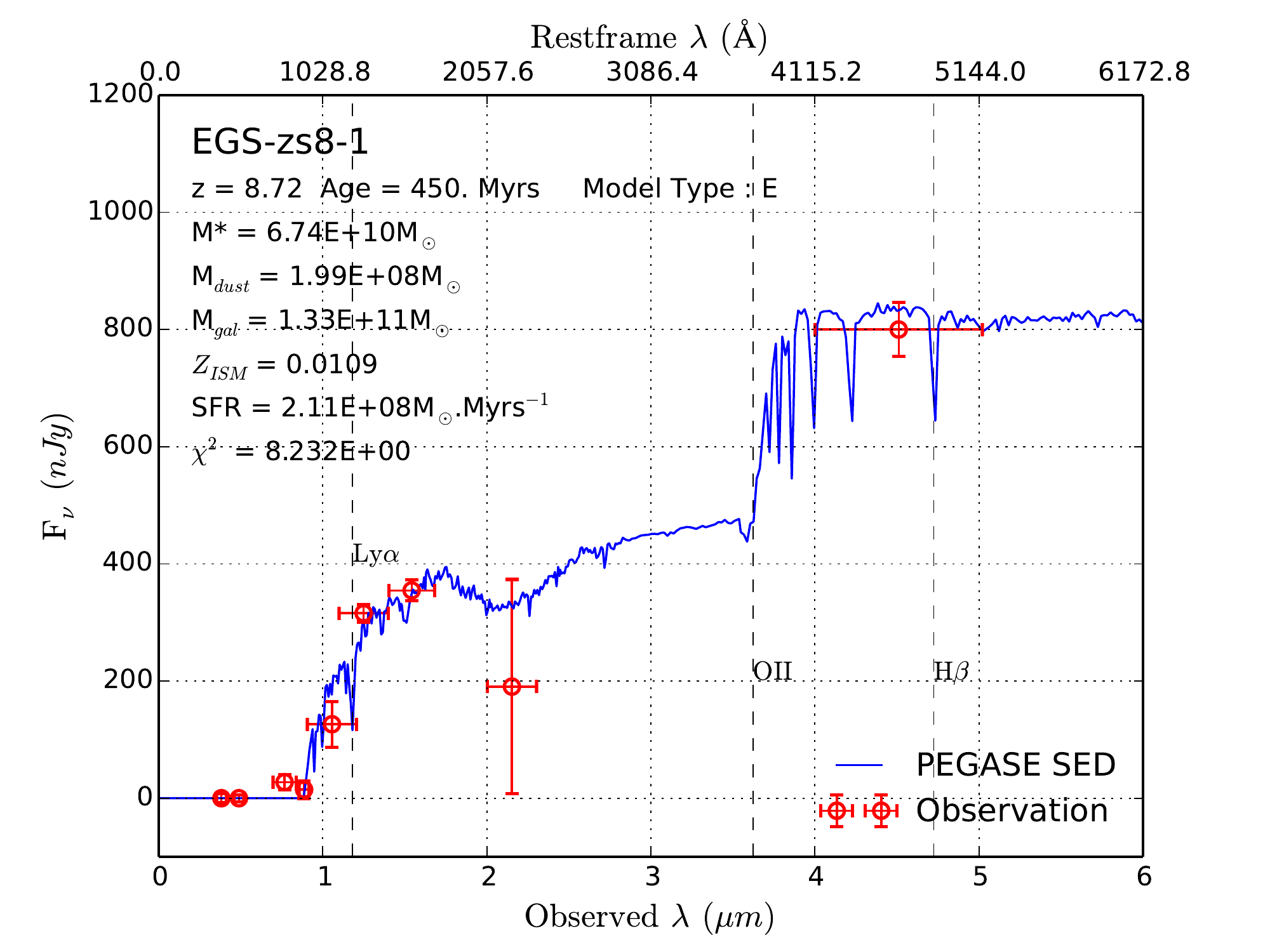}
\includegraphics[width=6cm,angle=0] {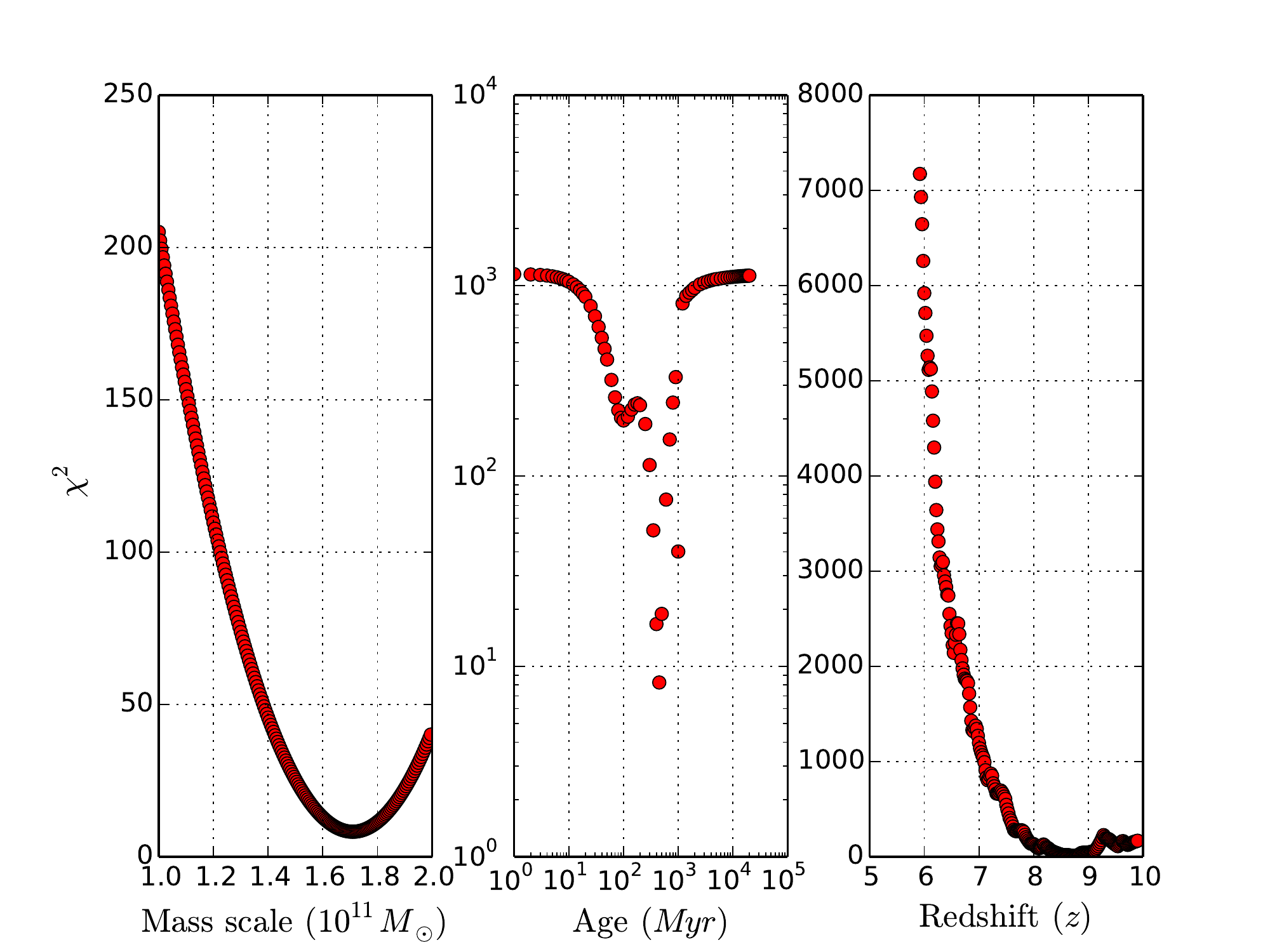}
\caption{\textit{Top: Best fit of of the galaxy EGS1 observations (red crosses) with the spheroidal SED shape (blue)  Bottom: its associated minimization curves. 
Main properties from the best-fitting  with the Pegase.3 spheroidal templates are given (see also Table~\ref{Results})}}
\label{EGS1E}
\end{center}
\end{figure}
\begin{figure}
\begin{center}
\includegraphics[width=12cm]{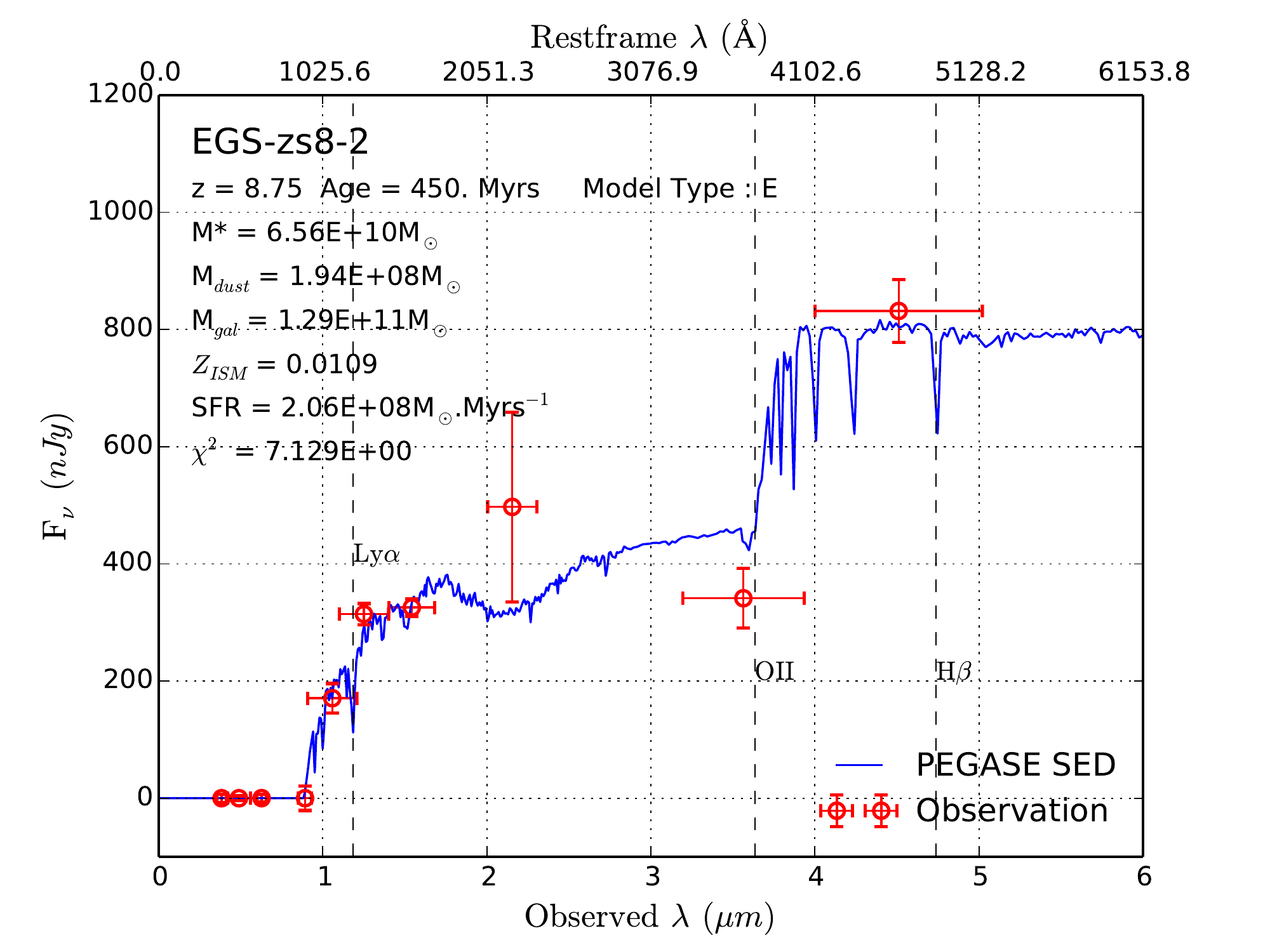}
\includegraphics[width=6cm,angle=0]{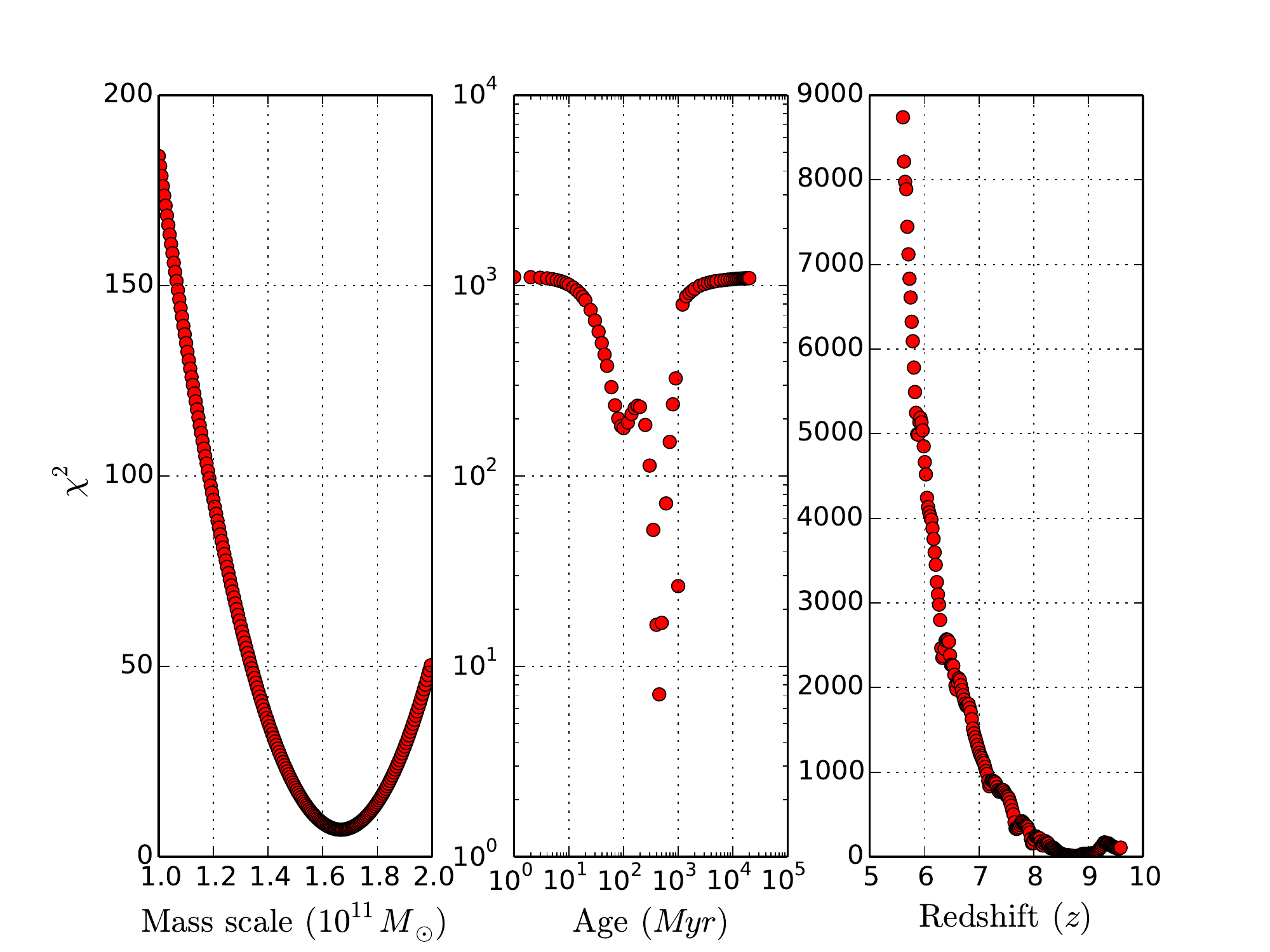}
\caption{\textit{Best fit of the spheroidal SED shape  with observations and associate minimization curves of the galaxy EGS2. Same comments as previous figures}} 
\label{EGS2E}
\end{center}
\end{figure}

Best-fits with error bars of the two other sample galaxies are presented  on 
Fig~\ref{EGS1E} and EGS2 on Fig~\ref{EGS2E}. 
Most other outputs as metallicity, masses after calibration are on Table~\ref{Results}. For EGS1, the best-fitted SED shape gives 
the photometric redshift $z_{LyB}$=8.72$^{+1}_{-1.32}$, within a few percent errors, consistent with the previously photometric redshift $z_{phot}$= 7.42. After tests with multiple libraries  of the two  (spheroidal, spiral ) geometries and slab, the best-fitted SED shape of the spheroidal geometry associated with the scenario SFH of elliptical galaxies (E) is  once more preferred. Geometry and SFH of spirals are not compatible with data while SB scenario is discussed in the last section. For EGS2, the best-fitted SED shape within a few $\%$\  gives the photometric redshift $z_{LyB}$=8.75$^{+1.25}_{-0.95}$. All other properties given on Table~\ref{Results} show the similarity of these two galaxies.
\\

\subsection{The case of the galaxy GN11}
At the present time, the most distant galaxy is GN-z11 at $z\simeq$11, \citet{Oesch2016}. The photometric redshift will only be tentative because the observational wavelength coverage does not cover the Balmer break. Fig~\ref{GN-z11} proposes the best-fit of observations where  the signature of Lyman  discontinuity and the attenuated continuum are well identified.  Once more while the spheroidal model best-fits data on the global wavelength coverage, the redshift determination is less robust  since observations are not covering the Balmer discontinuity. However the present results confirm a younger age of 160Myrs, a significantly less massive galaxy with reduced stellar mass and dust mass. These results, corresponding to the same SFH as adopted for the galaxy z=8 sample, are: 
 \begin{figure}
\begin{center}
\includegraphics[width=12cm,angle=0]  {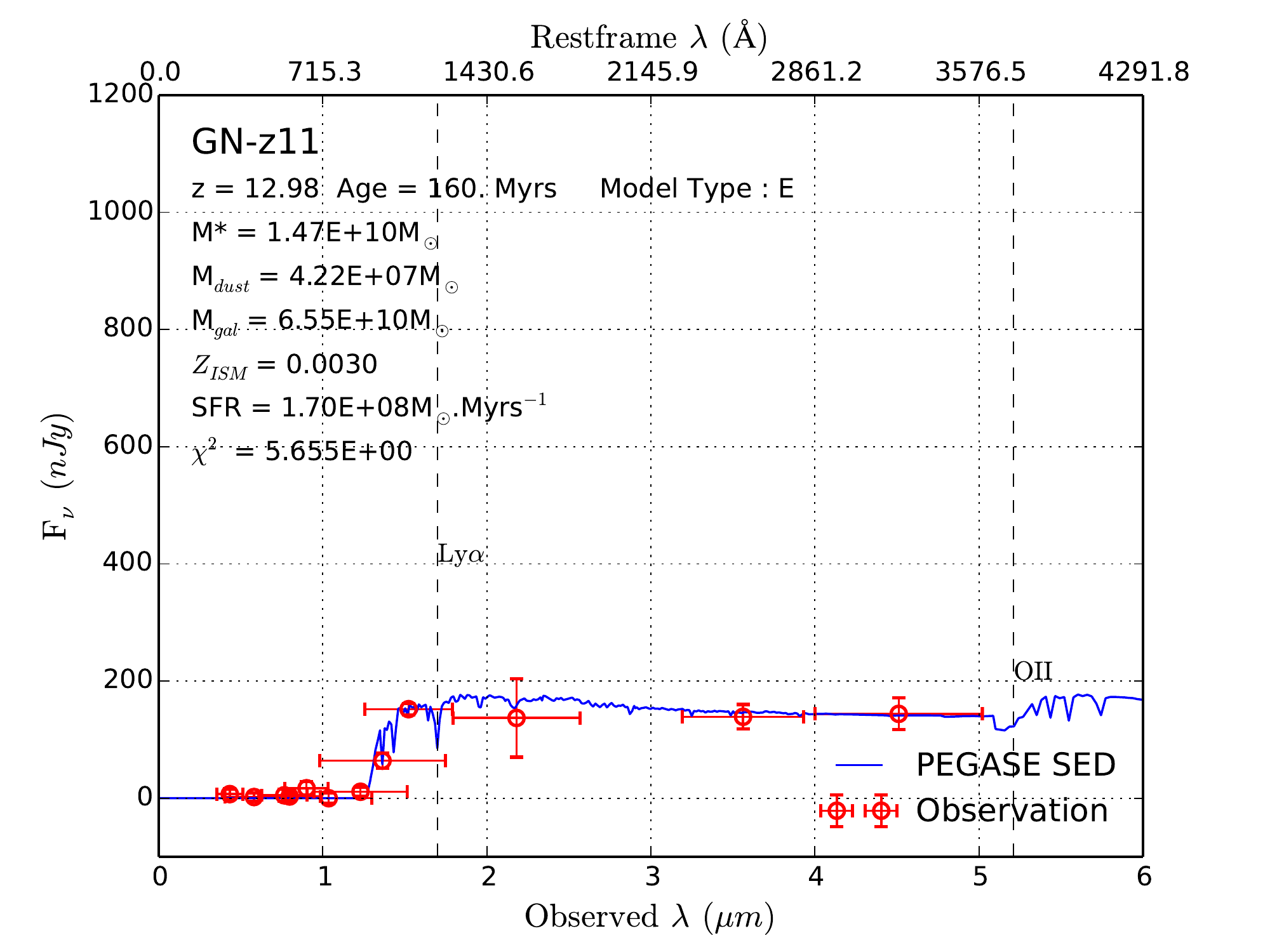}
\includegraphics[width=6cm,angle=0]   {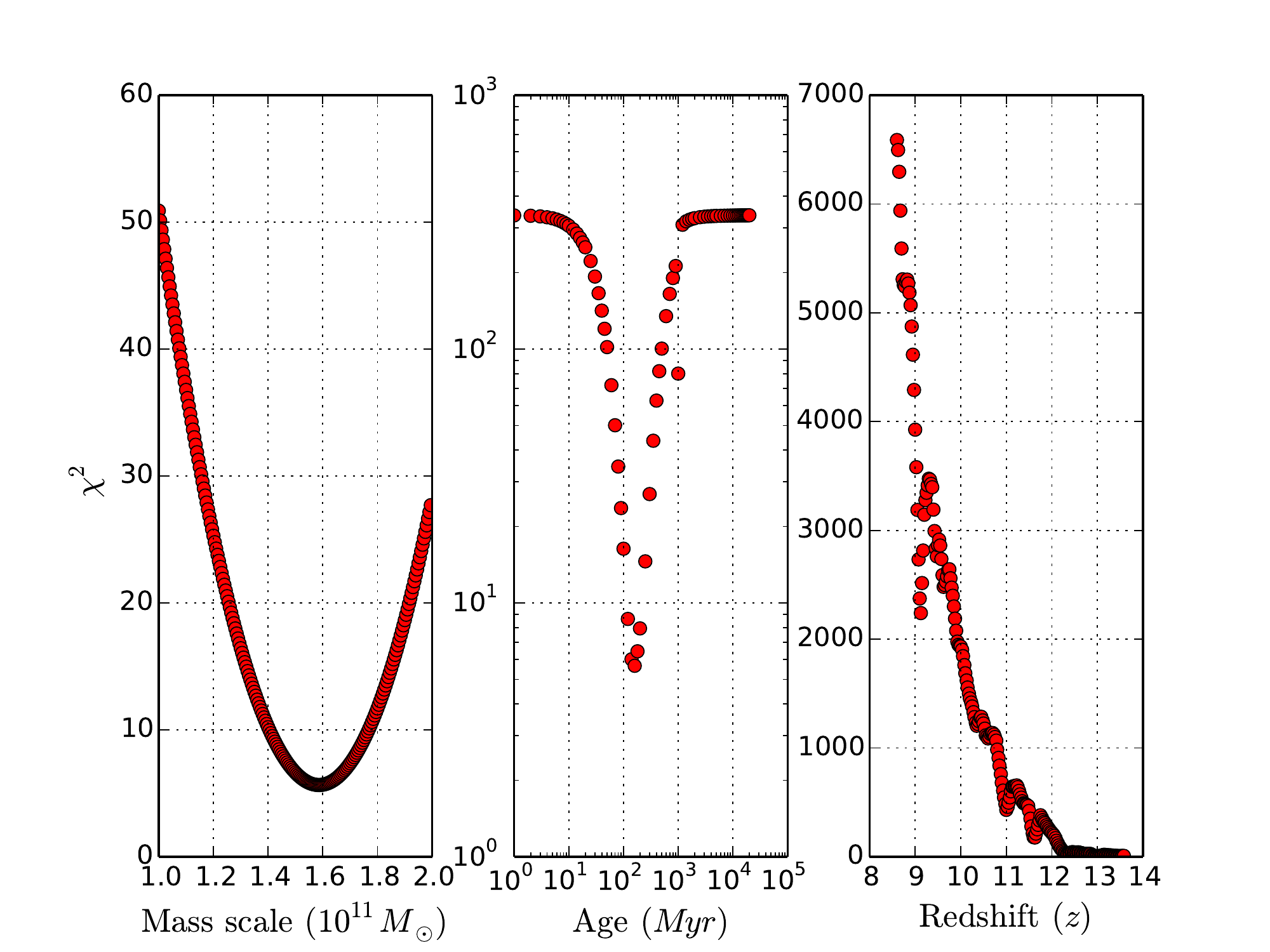}
\caption{\textit{Best fit of the spheroidal SED shape  with observations and associate minimization curves of the galaxy GN-z11. Redshifts less robust since the wavelength domain does not cover the Balmer discontinuity are not presented}} 
\label{GN-z11}
\end{center}
\end{figure}
\\
 \subsection {Summary of properties} 
 Table~\ref{Results} gathers the properties of our sample and GN-z11 from $\chi^2$ minimization procedure for the same  scenario of spheroidal galaxies. Its input parameters (SFH, inflow and outflow), also fitting the  z=0 elliptical  SEDs,  are:  
 
  $\psi(t)=\frac{(M_{ISM}(t))^a}{\tau_1}$;   $\dot{M}_{in} (t)=\frac{1}{\tau_2} e^{-t/ \tau_2}$ with $\alpha$=1, $\tau_1$ and $\tau_2$ = 300 Myrs and a maximal outflow at $t=1$ Gyr . 
 We recall that the spiral scenario, reproducing the local spiral (disk + bulge) SED, is not found compatible with the $z=8$ data and the instantaneous starburst is presented and discussed below.
 \\ 
\begin{table}
\begin{footnotesize}
\begin{center}
\begin{tabular}{c|c|c|c|c|c|c|c}
  Galaxy & age &M-gal & M-star & M-dust & Z & SFR (t)  & $\chi^2_{min}$\\
  \hline
  EGSY-2008532660& 500 & 12.5 & 6.8 & 0.0196 & 0.012& 182 &2.9\\
  \hline
  EGS-zs8-1 & 450  & 13.3 & 6.7 & 0.02. & 0.01 & 211& 8.2\\ 
  \hline
  EGS-zs8-2 &450& 13. & 6.6 & 0.02 & 0.01& 207 & 7.2\\
   \hline
  GN-z11 &  160 & 6.5 & 1.47 & 0.004 & 0.003& 170 & 5.6\\
 \end{tabular}
 \end{center}
\end{footnotesize}
\caption{\textit{Main properties of the sample and the GN-z11 galaxy: age of the stellar population in Myrs, masses in $10^{10}$ M$_{\odot}$, metallicity Z, SFR (t)  in $M_{\odot}$ yr$^{-1}$ from the best-fits by a $\chi^2$ minimization procedure.}}
\label{Results}
\end{table}

\section{Discussion  and conclusion}
With the help of multiple SED libraries of the new evolutionary farUV-to-farIR code Pegase.3, with coherent dust attenuation to emission modeling (Fioc and Rocca-Volmerange, 2018, submitted), we photometrically confirm the z$\simeq$8 redshifts of our selected sample from the two Lyman and Balmer break wavelengths. However we do not confirm the previous interpretations of an intense starburst inducing huge equivalent widths EW ([O III] and H$\beta$) of 1500$\AA$ based on no evolving models, without metal enrichment effects on the Ly$\alpha$ resonance line. 
As a main result our SED best fitting procedure attributes the typical [3.6]-[4.5]$\mu$m IR excess of the sample to the Balmer break induced by cumulated spectra of a red giant stellar population, the robust signature of an evolved elliptical galaxy, see \citet{Bruzual1983}. The difference with the previous  interpretation \citet{RobertsBorsani2016} is the SFH of an elliptical scenario for a MonteCarlo spheroidal geometry replaces the intense starburst. We show the predicted SEDs are in excellent agreement with observations on the whole UV-IR wavelength domain. The consequences are noticeable on ages and high stellar masses, star formation rates and SFHs, inflows and outflows. The fact that spiral scenarios and geometry simulations don't predict SEDs is no favoring disk galaxies. These $z=8$  galaxies could be primitive early-type galaxies, representative of local massive elliptical  galaxies.
As proposed by \citet{Decarli2017}, these galaxies of intense star formation rates but not embedding supermassive black holes (the galaxy sample is not known as active) could be close companions of quasars and maybe significant targets to clarify the  galaxy-AGN link.  
\begin{figure}
\begin{center}
\includegraphics[width=6cm,angle=0]{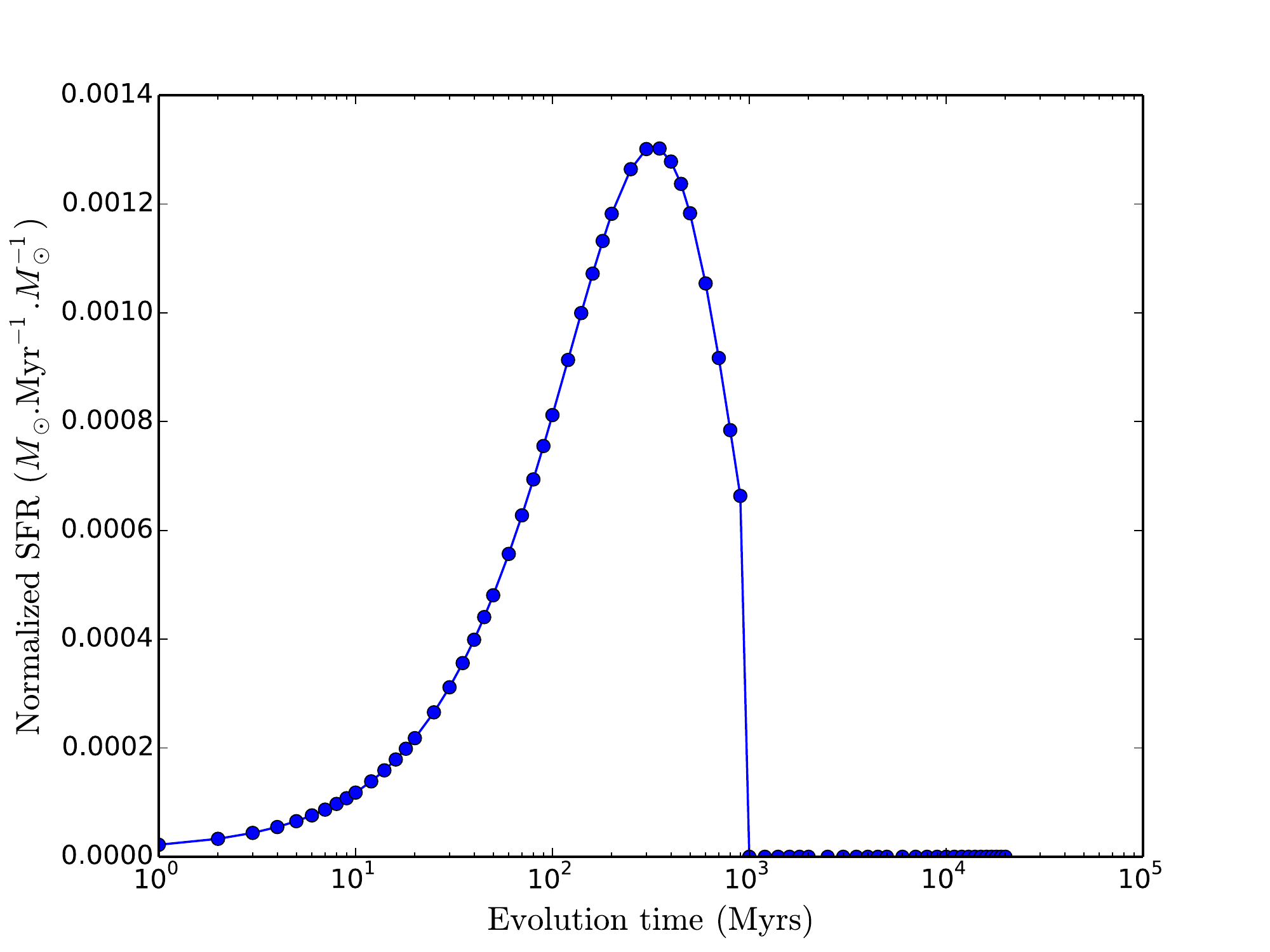}
\caption{\textit{The star formation history of spheroidal galaxies as a function of evolutionary time and  fitting the local color-color properties of ellipticals in the SDSS color diagram by spectral types see \citet{Tsalmantza2012}  }} 
\label{SFRz}
\end{center}
\end{figure}

\begin{figure}
\begin{center}
\includegraphics[width=16cm,angle=0]{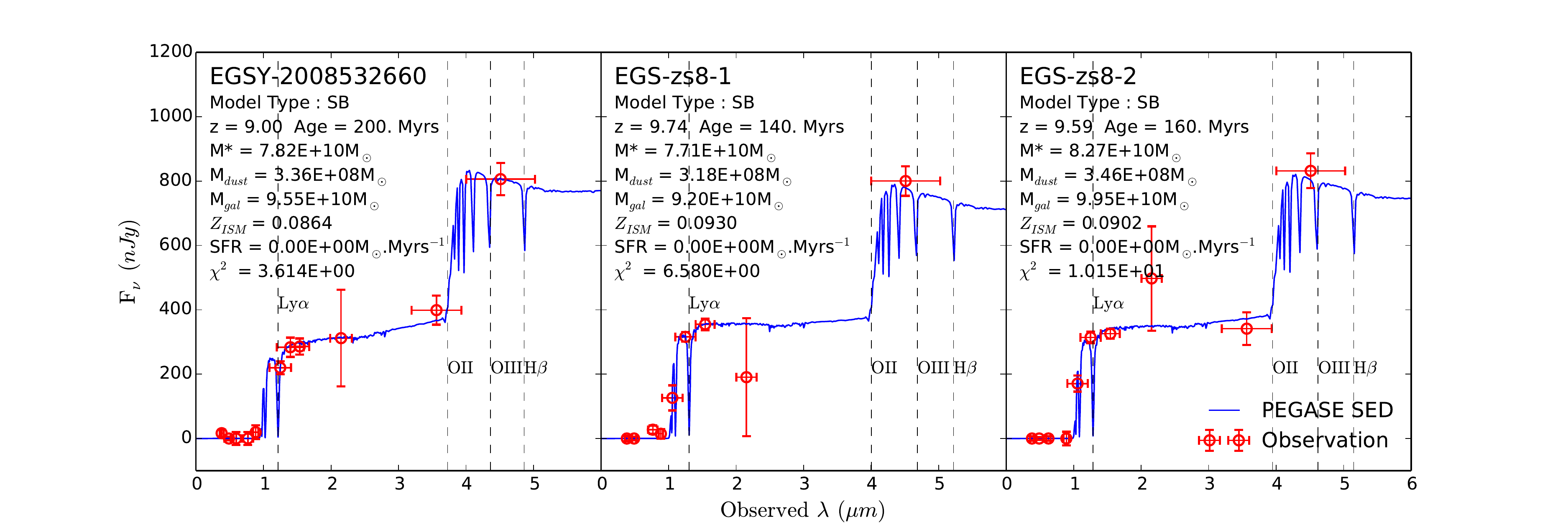}
\includegraphics[width=13cm,angle=0]{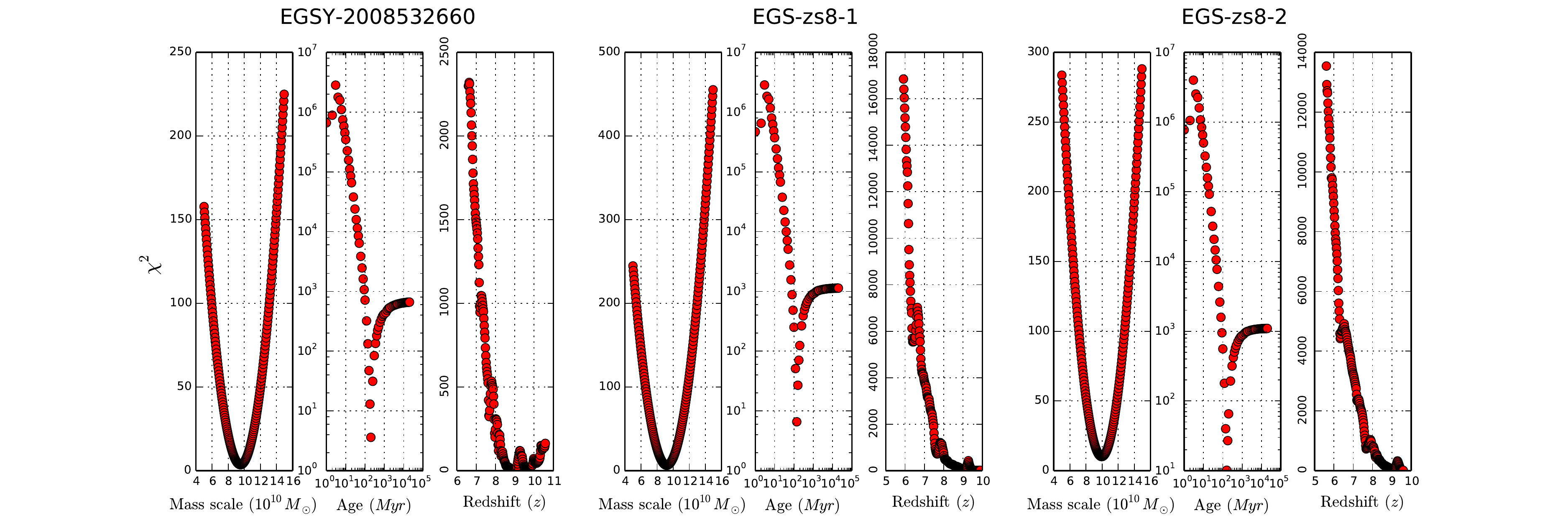}
\caption{\textit{Best fits of the galaxy sample with the less plausible scenario of instantaneous starburst (SB). Ages are significantly reduced but still massive galaxies of $10^{11}M_{\odot}$ formed.
}} 
\label{SBtemplates}
\end{center}
\end{figure}

The present results agree with the predictions of local elliptical/SO galaxies, as shown by the fitting of $0\leq z \leq 1$ SDSS observations  in a color-color diagram \citet{Tsalmantza2012}. Moreover as found by \citet{RoccaVolmerange2004}, these results confirm the interpretation of the most distant galaxies explaining the so-called $K-z$ distribution with models of highly massive ellipticals. The fact that models of spirals are unable to reproduce observations of these sources, likely no disks.  At such high redshifts the low surface brightness of disks likely prevents detection. Considering the instantaneous starburst (SB of 1Myr duration), such a  model is more or less academic, because it presents some difficulties to dynamically explain  a so rapid formation of $10^{11}M_{\odot}$  galaxies. However, our poor knowledge of the elliptical physics deserves to present the best-fits resulting of the SB scenario on Fig ~\ref{SBtemplates}. Ages vary from 140 to 200 Myrs and masses from 9.5 to 10$\times$ 10$^{10}$ M$_{\odot}$, quite coherent with results with E scenarios.

The interesting point concerns the spectral resolution of the observations defined by the filter passbands of the instrument. These interpretations in fact substitute the progressive attenuated slope to the brutal discontinuity, by age and dust attenuation effects. In these interpretations the equivalent widths of lines  are faint or null, in particular the resonance line Ly$\alpha$ line, disappearing with metal enrichment. In the revised version of P\'egase.3, in process of acceptation by the referee committee, the ionized emission lines will be computed as a function of metallicity, better adapted to the earliest phases of chemical evolution.  In such scenario, the emission lines ( [O III] and H$\beta$) exist but are very faint while  Ly$\alpha$ would disappear. We report to a further article, the computation of line equivalent widths with continua and emission lines, both depending on metallicity effects. 

In future analyzes, we will present the cumulated mass of stellar black holes from CCSN explosions after subtraction of ejecta, used as inputs for metal-enrichment, \citet{RoccaVolmerange2015}. The physical evolution of distant radio galaxies in relation with their galaxy host embedding a supermassive black hole is still a subject of active research \citet{RoccaVolmerange2018}.  The galaxy-AGN link  is also analyzed with the highest spatial resolution of the Gaia satellite. The redshift of formation extended to z=45 only increase the galaxy age by a few Myrs, according to the relation $z-t(z)$ with the previously cited cosmology parameters, not significantly changing our main conclusions. Discussions on the IMF evolution in primeval galaxies will also be considered, using the variety of IMFs proposed in the readme of P\'egase.3. Many fruitful results are waited for from the future satellites as JWST, ELT and EUCLID,  for their large wavelength coverage extended to the infrared and for the quality of their instruments.   
\bibliographystyle{aa}
\bibliography{references}
 
\end{document}